\documentclass[manuscript]{aastex}
\usepackage{spr-astr-addons}

\shorttitle{Oscillatory Universe}

\shortauthors{Ghosh et al.}

\begin{document}

\title{Oscillatory Universe, Dark Energy and General Relativity}

\author{Partha Pratim Ghosh\altaffilmark{1}}
\altaffiltext{1}{Department of Physics, A. J. C. Bose Polytechnic,
Berachampa, North 24 Parganas, Devalaya 743 424, West Bengal,
India\\ parthapapai@gmail.com}

\author{Saibal Ray\altaffilmark{2}}
\altaffiltext{2}{Department of Physics, Government College of
Engineering \& Ceramic Technology, Kolkata 700 010, West Bengal,
India\\ saibal@iucaa.ernet.in}

\author{A. A. Usmani\altaffilmark{3}}
\altaffiltext{3}{Department of Physics, Aligarh Muslim University,
Aligarh 202002, Uttar Pradesh, India\\ anisul@iucaa.ernet.in}

\and

\author{Utpal Mukhopadhyay\altaffilmark{4}}
\altaffiltext{4}{Satyabharati Vidyapith, North 24 Parganas,
Kolkata 700 126, West Bengal, India\\ utpalsbv@gmail.com}

\begin{abstract}
The concept of oscillatory Universe appears to be realistic and
buried in the dynamic dark energy equation of state. We explore
its evolutionary history under the framework of general
relativity. We observe that oscillations do not go unnoticed with
such an equation of state and that their effects persist later on
in cosmic evolution. The `classical' general relativity seems to
retain the past history of oscillatory Universe in the form of
increasing scale factor as the classical thermodynamics retains
this history in the form of increasing cosmological entropy.
\end{abstract}


\section{Introduction}
Although, the belief in Cyclic model related to {\it Oscillatory
Universe} dates back to the ancient times (\citet{Kanekar2001} and
references cited therein), a scientific model for it could only be
proposed during the first half of the twentieth
century~\citep{Friedman1922}. At a stage, cosmological model of
oscillatory Universe that combines both the Big Bang and the Big
Crunch as part of a cyclical event, was considered as one of the
main possibilities of cosmic evolution~\citep{Dicke1965}.
However,~\citet{Durrer1996} showed it as a viable alternative to
inflation. Throughout the past century, the idea has been of
scientific importance rather than of mere belief, which has been a
focus of many theoretical investigations till date as summarized
in the Section 2.

Various observations
\citep{Riess1998,Perlmutter1998,Spergel2003,Tegmark2004} suggest
that our Universe has been accelerating from $7$ Gyr ago. The
reason behind this phenomenon is believed to be a mysterious {\it
dark energy}, a term coined recently in 1999 but its history
traces back as late as Newton's time \citep{Calder2008}. The
cosmological constant $\Lambda$, first adopted and then abandoned
by Einstein for his static model of the Universe
\cite{Einstein1917,Einstein1918,Einstein1931}, is considered as
one of the candidates for the dark energy. It is, however,
speculated that $\Lambda$ is a dynamical term rather than a
constant. The other candidate is a scalar field often referred to
as quintessence \citep{Watterich1988,Ratra1988,Caldwell1998}.

The dark energy may be described as a perfect fluid through the
equation of state (EOS) relating the fluid pressure and matter
density of the physical system through the relation
\begin{equation}
\label{eq1} p=\omega\rho,
\end{equation}

where  $\omega \equiv \omega(t)$ is the barotropic dark energy EOS
parameter. It plays a significant role in the cosmological
evolution. This is in general a function of time and may be a
function of scale factor or redshift
\citep{Chevron2000,Zhuravlev2001,Peebles2003}. Recently,
\citet{Usmani2008} have proposed a time dependent $\omega$ for the
study of cosmic evolution with the EOS parameter
\begin{equation}
\label{eq1} \omega(t)= \omega_0 + \omega_1 (t \dot H/H).
\end{equation}

Here, $H\equiv H(t)$ is the time varying  Hubble parameter. The
dynamics of both expanding and contracting epoch of the Universe
is buried in the sign of $\dot H$ in this simple expression. The
$\dot H$=0 represents a linearly expanding Universe with
$\omega(t)=\omega_0$.

A close system like an oscillatory Universe may be created without
violating any known conservation law of Physics as its total mass
(energy) is zero, which is indeed possible \citep{Tryon1973}. A
Universe may enter into an epoch of expansion followed by an epoch
of contraction after every bounce. For an expanding Universe that
was created at $t=0$ with $H(t)=0$, we may arrive at two
situations: (i) linearly expanding Universe represented by
$H=t\dot H$, and (ii) Universe that has undergone an adiabatic
expansion to a finite $H_i$ followed by a linear expansion (for
example, our Universe) that may be represented by another
condition $ H >>t\dot H$ provided $H_i >> t\dot H$. Having the
same theoretical framework these two cases may be represented by a
single set of equations \citep{Usmani2008} with a difference that
inflation leads to a scaling in $\omega$. Thus, it is the EOS that
distinguishes between the two. After achieving  its maximum radius
at time $t=t_0$ ($\tau =0$) with $\dot H=0$ and $H(t_0)=H_{max}$,
the Universe starts collapsing due to its own mass and  enters
into an epoch of contraction with a negative sign of $\dot H$.
Thus at a later time $t=t_0+\tau$, where $H$ is defined as
$H=H_{max} -\tau\dot H$, which would continue to decrease till
Universe achieves its  minimum radius. At preceding times closer
to it, a `condition of oscillation', $ H(t)<< |t\dot H|$, would
hold. This condition would still at a time when Universe sets to
expand again.

The above EOS enables us to handle this third possibility at least
qualitatively when employed in the framework of general
relativity. The history of oscillations does not go unnoticed by
the classical thermodynamics since cosmological entropy
(unidirectional thermodynamical scale of time) keeps on growing
after every cycle \citep{Tolman1934}. The generalized second law
of thermodynamics relates the dark energy EOS with cosmological
entropy
\citep{Bakenstein1973,Gibbons1977a,Gibbons1977b,Unruh1982,Davies1988}.
Thus the EOS, which translates itself into entropy, would manifest
similar properties despite the fact that relativity breaks down at
singularities, whose details are not required in the framework. An
observer outside the Universe, who measures a growing time (or
growing entropy) from point of creation, would distinguish a state
in an oscillation with an identical state in the next oscillation
with a decrease in  $\dot H$. This would result in different
solutions of field equations for identical states in two different
oscillations. The state of expansion, contraction and oscillation
of the Universe is buried in  $\dot H$. We find no reason to
believe, why field equations of general relativity would not yield
a valid solution for any value of $\dot H$, no matter how many
times the Universe has hit the singularity in the middle and has
come up to a finite size again and again for `unknown' reasons.

\section{A brief overview} The building up of cosmological
entropy in each cycle of oscillation as shown by
\citet{Tolman1934} is fated to a thermodynamical end. This
suggests that the Universe had a beginning at a finite time, and
thus had undergone a finite number of cycles
\citep{Zeldovich1983}, opposed to the idea of steady-state
Universe. When treated classically, it reaches a point of
singularity at the end of every cycle resulting in a total
breakdown for the general relativity (GR). However, there are
models with contracting epoch preceding a bounce in M-theories
\citep{Khoury2002,Steinhardt2002}, braneworlds
\citep{Kanekar2001,Shtanov2003,Hovdebo2003,Foffa2003,Burgess2004}
and loop quantum cosmologies \citep{Lidsey2004,Ashtekar2006}. The
quantum effects built into these studies aim to provide a
non-singular framework for the bounces that avoid a kind of
singularity hit by GR. Besides, they all admit a cosmological
evolution that has undergone an early oscillatory phase for a
finite time, and has finally led to inflation as we see it now.
\citet{Saha2004} obtained the oscillatory mode of expansion of the
Universe from a self-consistent system of spinor, scalar and
gravitational fields in presence of a perfect fluid and
cosmological term $\Lambda$ \citep{Zeldovich1967,Weinberg1989}.
Evolution of the expanding Universe has also been studied by using
scalar, vector and tensor cosmological perturbation theories
\citep{Mukhanov1992,Noh2004,Bartolo2004,Nakamura2007}. Scalar
field models that involve contracting phases are proposed as
alternatives to the inflationary scenario. These models have also
been explored to distinguish such phases from purely expanding
cosmologies. Vector perturbations uptill the second order have
recently been considered as the scalar perturbations may not be
able to distinguish it at first order \citep{Mena2007}.
Oscillatory model arising from linearized $R^2$ theory of gravity
\citep{Corda2008} has also been studied, which is found in
reasonable agreement with some observational results like the
cosmological red shift and the Hubble law.

Besides the above works on oscillating Universe models we would
like to have a special mention of the following two models:\\ (i)
The Steinhardt-Turok model which gives a description of two
parallel M-branes that collide periodically in a higher
dimensional space \citep{Steinhardt2005}. Here dark energy
corresponds to a force between the branes.\\ (ii) In the
Baum-Frampton model EOS parameter assumes the form $\omega < -1$,
which is known as phantom energy condition in contrast to the
Steinhardt-Turok assumption where $\omega$ is never less than $-1$
\citep{Baum2007}.

These two models will provide motivation for choice of the form of
EOS parameter in connection to our model on oscillatory Universe.

\section{The Einstein field equations and their solutions}
The Einstein field equations of GR can be given by
\begin{eqnarray}
\label{eq2} R^{ij}-\frac{1}{2}Rg^{ij}= -8\pi
G\left[T^{ij}-\frac{\Lambda}{8\pi G}g^{ij}\right],
\end{eqnarray}
where $\Lambda$ is time-dependent Cosmological constant with
velocity of light in vacua being unity (in relativistic units).
The justification for introducing this time varying $\Lambda$ can
be found directly from the observations on supernova due to the
High-$z$ Supernova Search Team (HZT) and the Supernova Cosmology
Project (SCP) \citep{Riess1998,Perlmutter1998} as discussed in the
Introduction. To understand the physical nature of this exotic
energy that tends to increase the speed of expansion of the
space-time of the Universe several models have been proposed by
the investigators as can be found in the literature
\citep{Overduin1998,Sahni2000}.

For the spherically symmetric
Friedmann-Lema{\^i}tre-Robertson-Walker metric, the Einstein field
equations (\ref{eq2}) yield Friedmann and Raychaudhuri equations,
respectively, as follows
\begin{eqnarray}
\label{eq3} 3H^2+\frac{3k}{a^2} &=& 8\pi G\rho+\Lambda,\\
\label{eq4} 3H^2+3\dot H &=& -4\pi G(\rho+3p)+\Lambda.
\end{eqnarray}

Here, $a=a(t)$ is the cosmic scale factor, $k$ is the  curvature
constant, $H=\dot a/a$ is the Hubble parameter and $G$, $\rho$,
$p$ are the gravitational constant, matter-energy density and
fluid pressure, respectively. The $G$ is taken to be a constant
quantity along with a variable $\Lambda$ where the generalized
energy conservation law may be derived \citep{Shapiro2005}. Using
the EOS (1) in Eq.~(\ref{eq4}) and comparing it with
Eq.~(\ref{eq3}) for flat Universe ($k=0$), we arrive at
\begin{equation}
\label{eq5} \dot H = -4\pi G\rho (1+\omega).
\end{equation}

In the light of the studies by \citet{Ray2007}, we use the {\it
ansatz} $\label{eq6} \dot\Lambda= AH^3$, where $A$ is a constant
of proportionality. As argued by \citet{Usmani2008} this {\it
ansatz} may find realization in the framework of self consistent
inflation model \citep{Dymnikova2000} in which time-dependent
$\Lambda$ is determined by the rate of Bose condensate evaporation
with $A \sim (m_B/m_P)^2$, where $m_B$ is the mass of bosons and
$m_{P}$ is the Planck mass.

The EOS along with Eqs.~(\ref{eq4}),~(\ref{eq5}) and above {\it
ansatz} leads to
\begin{equation}
\label{eq7} \frac{d\dot H}{dH}+3(1+\omega)H=
\frac{A(1+\omega)H^3}{2\dot H},
\end{equation}
with $dH/dt=\dot H$.

The condition $ H(t)<< |t\dot H|$ with $\pm \dot H$ simplifies it
further as
\begin{equation}
\label{eq8}  \frac{d\dot H}{dH} \pm 3\omega_1 \tau \dot H=0.
\end{equation}

A change of sign for $\dot H$ would amount to the  change  of sign
for  $\omega_1$. Thus, we have same solution set for $\pm \dot H$
with $\pm \omega_1$ as follows

\begin{eqnarray}
\label{eq15} a(t)&=&C~{\rm exp}(B[ln~T-1]),
\end{eqnarray}

\begin{eqnarray}
\label{eq16} H(t)&=& \frac{B}{t} ln~T,
\end{eqnarray}

\begin{eqnarray}
\label{eq17} \omega (t)&=& \omega_0+\omega_1
\left(\frac{1}{ln~T}\right),
\end{eqnarray}

\begin{eqnarray}
\label{eq18} \rho (t)&=& - \left(\frac{1}{4\pi G}\right)
                         \left[\frac{B}{ t^2 (1+\omega(t))}\right],
\end{eqnarray}

\begin{eqnarray}
\label{eq19} p (t)&=& \omega (t) \rho (t),
\end{eqnarray}

\begin{eqnarray}
\label{eq20} \Lambda (t)&=& \frac{A B^3}{t^2}
\left[(ln~T)^3-3(ln~T)^2+6~ln~T-6\right],
\end{eqnarray}

with $B=t/E\tau$, $T=DE\tau t$ and $E=3\omega_1$, whereas $C$ and
$D$ are two constants of integration.

\section{Physical features of the oscillatory model}
The term $ln~T$ demands a positive definite value for $T$. Thus
$E$, $D$ and $\omega_1$ must have identical sign. Therefore,
second term of  Eq.~(\ref{eq17}) i.e. $\omega_1/ln~T$ is positive
for $T > 1$ and negative in the range  $0< T <1$. We find
$\omega_1=0$ for a linearly expanding Universe and a range $-2/3
<\omega_1<-0.46$ for an inflationary Universe \citep{Usmani2008}.
Thus, it is always negative. For a contracting Universe $\dot H$
is negative, which makes the second term of Eq.~(\ref{eq2})
positive. This, when compared with Eq.~(\ref{eq17}), suggests that
$\omega_1$ is positive for  $T>1$ and negative for $T<1$. Thus,
for $D=1$, we obtain a condition, $ E=3\omega_1  >  1/ \tau t $.
Here $t \rightarrow \infty$ means $\omega_1 \rightarrow 0$, which
represents a linearly expanding Universe with
$\omega(t)=\omega_0=-1/3$. For an observer sitting outside the
Universe, the rate $\dot H$ would appear to be smaller and smaller
for the increasing number of oscillations  as $t$ would appear to
be larger and larger. It is obvious from Eq.~(\ref{eq16}) that $H$
is positive definite that means it grows from zero to a maximum
value during the epoch of expansion and and then approaches to
zero during the epoch of contraction. We also observe that
Eq.~(\ref{eq18}) for the density is singular at $1+\omega(t)=0$,
so is the pressure ($p=\omega\rho$), which turns out to be
positive for $0>\omega(t)>-1$. The same restriction is found for
the expanding Universe using same EOS \citep{Usmani2008}, which is
in agreement with the results of GSL of thermodynamics that too
restricts the EOS, $\omega(t) >-1$, in an expanding Universe.
Beyond this limit, for example $\omega(t)>0$, pressure is
negative.

Let us now try to extract the two phases of expanding and
contracting Universe. As mentioned earlier, we had the condition
$-1 < \omega_1 <-0.46$ \citep{Usmani2008} which we must satisfy
here also and thereby take $\omega_0 =-1/3$ and $-2/3 < \omega_1 <
-0.46$. With this our $\rho$ will always be positive either
Universe is expanding or contracting. In our solution $T$ depends
upon the constants $D$, $E$, $\tau$ and also $t$. $E$, via the
relation $E = 3 \omega_1$, is a negative quantity. Thus $D$ must
be negative to make $T$ positive and between $0$ and $1$. Hence
pressure will be positive and Universe will contract. It also
depends upon constant $D$, so if $|D|$ multiplied by $\tau$ is
less than $1$ pressure would be positive. In other words if $E$ is
very very small, even for a large $\tau$, pressure will be
positive and Universe will contract. But with increasing $\tau$,
$|D|\tau$ will eventually be greater than equal to one and
pressure will again be negative and the Universe will start
expanding.

In the Fig. 1 we have depicted the feature of expansion and
contraction via the scale factor $a(t)$. It can be observed from
the figure that growth of scale factor $a(t)$ for the successive
oscillations have the equal epoch of expansion and contraction.

\begin{figure}
\begin{center}
\vspace{0.5cm}
\includegraphics[scale=0.8,clip,keepaspectratio=true]{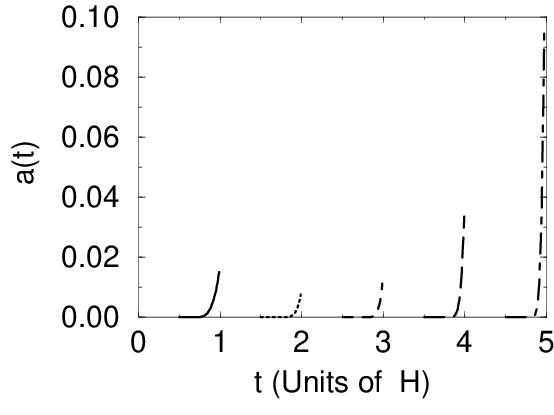}
\caption{Growth of scale factor $a(t)$ for the successive
oscillations having equal epoch of expansion and contraction. The
solid, dotted, dashed, long-dashed and chain curves represent
$a(t)$ during first, second, third, fourth and fifth oscillations,
respectively. The oscillation period for the Universe is assumed
to be ($t=1$ units of H). } \label{fig1}
\end{center}
\end{figure}

\section{Conclusion}
We have constructed in the present work the story for a Universe
like ours own. The Universe had started with an adiabatic
expansion with $\omega(t) <-1/3$, then it had been slowed down to
a linearly expanding Universe with $\omega (t)=-1/3$, which was
then followed by an epoch of contraction with a positive value of
$\omega_1/ln~T$. For smaller values of $T$ or $t$, $\omega_1/ln~T$
may reach the condition $\omega(t)>0$, which will make the
pressure positive and thus would receive a bounce. This process
would continue repeating without violating any known conservation
law. However, $\omega_1/ln~T \rightarrow 0$ for $T\rightarrow
\infty$. Thus, the Universe would appear to be a linearly
expanding Universe with $\omega(t)=\omega_0=-1/3$ and would never
reach the condition for contraction or bounce i.e. $\omega(t)>0$.
Thus, it recommends for initial oscillations only. One may suggest
improvements in the EOS, however we have shown a way to
incorporate oscillating Universe in the general relativity through
it.

Some other suggestions may also be correlated with the present
work, e.g. pertinent issues on the origin of the observed baryon
asymmetry and also physical nature of the dark matter. In
connection to the evolution of the baryon charge in the
oscillating baryon asymmetric Universe we would like to mention a
few recent works
\citep{Buchmullera2005,Davoudiasl2010,Canettia2012}, in particular
the review work of \citet{Canettia2012} where they discuss a
testable minimal particle physics model that simultaneously
explains the baryon asymmetry of the universe, neutrino
oscillations and dark matter. So, following the `mechanism for
generating both the baryon and dark matter densities of the
Universe' \citep{Davoudiasl2010} a future project may be
undertaken within the purview of our present investigation.

\acknowledgments SR and AAU are thankful to the authority of
Inter-University Centre for Astronomy and Astrophysics, Pune,
India for providing Visiting Associateship under which a part of
this work was carried out. We all are also thankful to the referee
for raising some pertinent issues which helped us to improve the
standard of the manuscript.

\end{document}